\documentclass[10pt ]{revtex4}
\usepackage{hyperref}
\usepackage{amsmath,amsfonts}
\usepackage[dvips]{graphics}
\usepackage[dvips]{graphicx}
\usepackage{subfigure}
\usepackage{epsfig}

\RequirePackage{color}

\begin{document}

\title{Large Distance Modification of Newtonian Potential  and  Structure Formation in Universe} 
\author{Mir Hameeda$^{1}$ 
} \email{hme123eda@gmail.com } 
\author{Sudhaker Upadhyay$^{2}$ } \email{sudhakerupadhyay@gmail.com}
 \author{ Mir Faizal$^{3}$} \email{mirfaizalmir@googlemail.com}
 \author{Ahmed F. Ali$^{4}$} \email{ahmed.ali@fsc.bu.edu.eg}
 \author{ Behnam Pourhassan$^{5}$ }\email{b.pourhassan@du.ac.ir}
 \affiliation{
$^{1}$Department of Physics, S.P. Collage,  Srinagar, Kashmir, 190001, India and \\
  Visiting Associate, IUCCA,  Pune,  41100, India}
 \affiliation{$^{2}$Department of Physics, K.L.S. College, Nawada, Nawada-805110, India}
 \affiliation{ $^{3}$Irving K. Barber School of Arts and Sciences, University of British Columbia - Okanagan,\\  3333 University Way, Kelowna,   British Columbia V1V 1V7, Canada and\\
  Department of Physics and Astronomy, University of Lethbridge, Lethbridge, Alberta, T1K 3M4, Canada}
 \affiliation{$^{4}$Department  of Physics, Faculty of Science,
 Benha University, Benha, 13518, Egypt}
 \affiliation{$^{5}$School of Physics, Damghan University, Damghan, 3671641167, Iran}

\begin{abstract}
In this paper, we study the effects of super-light brane world
perturbative  modes
on structure formation in our universe. As these modes modify the
large distance behavior of Newtonian potential, they effect the
clustering of a system of galaxies.
So, we explicitly calculate the clustering of galaxies interacting through such a
modified Newtonian potential. We  use  a suitable approximation for analyzing
this system of galaxies, and discuss the validity of such approximations. We
observe  that such  corrections also modify the  virial theorem
for such a system of galaxies.
\end{abstract}
 \maketitle

\section{Introduction}
We   approximate the galaxies as point particles,
and analyze the clustering of a system of such galaxies. This approximation will be
valid as the distance between two galaxies is many orders of magnitude larger than the
size of a single galaxy. Thus, we   use techniques of  standard statistical mechanical
to analyze the clustering of a system of galaxies. It may be noted that such an
analysis has already been performed using the usual Newtonian potential
\cite{sas84, ahm02,ahm06,ahm14, rup96}, and thus the techniques of  statistical
mechanics has already been used to analyze the clustering of galaxies.
However, we have to either consider dark matter, or a modified Newtonian
potential to explain the physics at large scales, we will analyze the clustering of
galaxies using a Newtonian potential modified by super-light modes of a brane world model.
We would like to point out that in this formalism a  cosmic energy equation for a system of
galaxies  was obtained   using the standard techniques of statistical  mechanics
\cite{sut90,sut93}.  This was used to analyze the   clustering of a system of galaxies
using correlation functions \cite{bla59, hu01, pee01}.  In this formalism,
the correlation function  and the power spectrum  characterize the distribution of
galaxies in  clusters and superclusters  \cite{pee80}.

So, in this paper, we   analyze
the clustering of galaxies using this formalism. In fact, we will use the large distance
corrections to the Newtonian potential from super-light modes in brane world models \cite{modes}.
These cosmological models have been motivated from
string theory due to extra dimensions in string theory  \cite{ran99}.
In these brane world models,  our universe is a brane in a higher dimensional bulk.
These models have been used for  resolving  the hierarchy problem and the weakness of
gravitational force in comparison with other three fundamental forces \cite{ark00}.
In fact, even though there are different models  for brane world theories \cite{bro06},
a  common feature of  all of these different models is that   the standard model fields are confined to the four dimensional brane and the  gravitons propagate into the bulk   confined to the brane and thus can propagate into the higher dimensional bulk
\cite{ark99, ark98, ant98}. Due to the propagation of gravity into higher dimensions,
the Newtonian potential gets brane corrections. Furthermore, as the general relativity
along with its Newtonian approximation have not been tested at very large or very small
distances, it is possible that  the Newtonian potential would get modified at such distances. Generally, Newtonian potential may
be modified due to several effects like dark energy \cite{1,2}. So, usually, the corrections generated from brane world gravity  modify the Newtonian potential at small distances \cite{flo99, ran99}, and these modifications cannot produce any new astrophysical  or cosmological effects.
However, it is possible to obtain  super-light perturbation modes in brane world models, and these super-light modes can  modify gravity at large scales  \cite{modes, brane12}.
The importance of the models with super-light   perturbation modes is due to the fact that  these
predict a modification of the gravitational interaction for matter on the brane at astronomical scales. The corrections to Newton's gravity due to such consideration may be promising for resolving the issue  of dark matter  in galaxies and galaxy clusters and even the cosmological dark energy problem \cite{modes}.
The form of corrected Newtonian potential is given as  \cite{modes1}
\begin{equation}
 \phi  = \phi_N \left( 1 + \frac{k}{r^2}\right),\label{ori}
\end{equation}
where $\phi_N$ is the standard Newtonian potential given by,
\begin{equation}
 \phi_N = - \frac{G m^2 }{r}.
\end{equation}
It may be noted that this long distance correction scales as $1/r^3$, which is
unlike  the short distance correction which scales as  $1/r^2$ \cite{modes}.
As this correction changes the Newtonian potential at large distances, this can be  used in analyzing the dynamics of galaxies \cite{gala1, gala2, gala4, gala5}.
In fact, it has been demonstrated that
the brane world models can explain the rotation curve of galaxies better than
the models which are based on the existence of  dark matter  \cite{matter}.

It may be pointed out that phenomenologically motivated modified theories of gravity (MOG) \cite{3,4}
have been used as an alternative to dark matter. In fact, modified Newtonian dynamics
(MOND) \cite{mil83} and MOG \cite{mof15, ros15}, as two possible modified theories of
gravity have been used to obtain the correct rotation curves of galaxies.
The MOG modifies the large distance behavior of  Newtonian potential  \cite{mof15, ros15},
and this modification produces the correct rotation curve of galaxies.
Thus, it is important to consider  large distance correction to Newtonian potential
for analyzing astrophysical phenomena. An advantage of using the corrections from
super-light perturbation modes in brane world models is that such corrections are
motivated from theoretical considerations and cosmological models motivated from string theory,
but they can also have interesting phenomenologically applications
\cite{gala1, gala2, gala4, gala5}. In this paper, we will use this long distance
correction to the Newtonian potential produced by super-light modes,
and analyze its effect on the clustering of galaxies.

Moreover, one may note that even though there are problems with certain  distance based modifications of gravity, such
as MOND, in order to explain the clustering of galaxies \cite{dod, str, chan}, it has been argued that other kind of modifications
to gravity can explain clustering of galaxies \cite{mof, hod}. In fact, it  is possible to modify MOND in such a way,
that  force law approximates MOND at large and intermediate accelerations, and gets further modified
at    ultra-low accelerations. Such ultra-low accelerations are  relevant to  the
galaxy clusters, and such a modification
has been observed to be consistent  with the observations \cite{hod1, zha, kho, mil}.
It has  been demonstrated that  MOG, which modifies the Newtonian law of gravitation, can consistently explain
the clustering of galaxies \cite{mar, mof0, bro}. It is  also possible to explain
the clustering of galaxies without dark matter by using  a  modified theory of gravity based on
covariant Galileon model  \cite{sal}.
So, even though the modification of gravity such as  MOND cannot be used
to analyze the clustering of galaxies, it is possible to have alternative theories of gravity, which may explain such a model.

So, even though there are problems with MOND in explaining clustering of galaxies, it is possible
to consider other models of modified  gravity, which do not have above discussed problems. Furthermore, as the clustering of galaxies
has been studied using techniques of statistical mechanics with Newtonian potential \cite{yan, ahm, yan1, leo, ahm1, hur},
it would be both important and interesting to generalize
such an analysis of modified law of Newtonian gravity. Even though there might still be problems with such an approach, it would
be a better approximation to explain the clustering of galaxies. We could improve this analysis further by incorporating
dark matter, but the use of modified Newtonian  potential would produce better results than the standard Newtonian potential.
It has been argued that the  modified theories of gravity are produced from dark matter models
\cite{bal, ber, ber1, kho1}.
It has also been discussed that  clustering
can be explained using brane world models \cite{gala2, har, chak, alc, hey}.
So, this motives us to use   techniques of statistical mechanics \cite{yan, ahm, yan1, leo, ahm1, hur},  with  brane world modified  Newtonian potential, to galactic clustering.

\section{Clustering Parameter} \label{sec:style}
\begin{figure}
\begin{center}$
\begin{array}{cccc}
\includegraphics[width=70 mm]{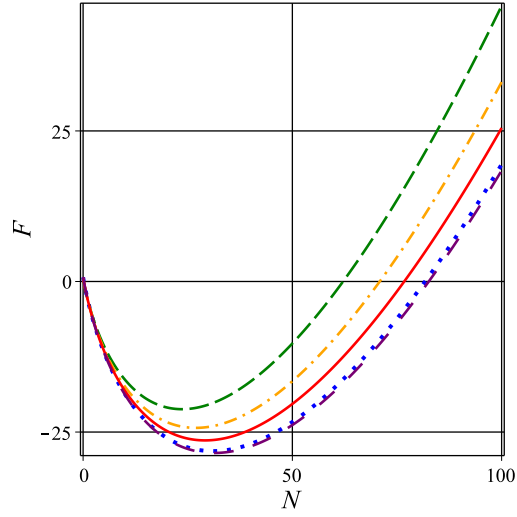}
\end{array}$
\end{center}
\caption{Typical behavior of Helmholtz free energy in terms of $N$. We set unit values for all parameters. Blue dotted line represents the case of $\epsilon=k_l=0$, green dashed line represents the case of $\epsilon=1$, $k_l=0$, orange dash dotted line represents the case of $\epsilon=1$, $k_l=0.6$, red solid line represents the case of $\epsilon=1$, $k_l=1$, violet space dash line represents the case of $\epsilon=1$, $k_l=1.4$.}
\label{fig0}
\end{figure}

\begin{figure}
\begin{center}$
\begin{array}{cccc}
\includegraphics[width=70 mm]{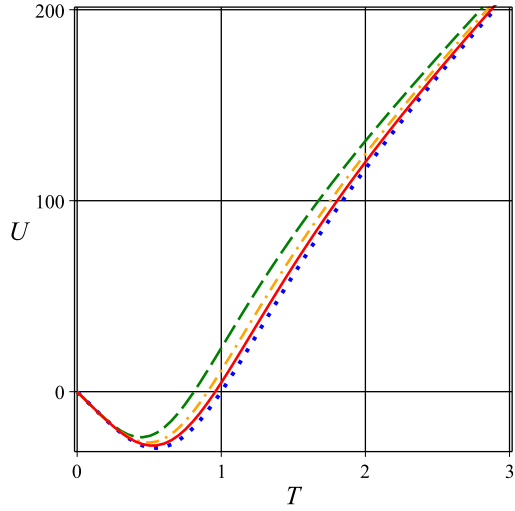}
\end{array}$
\end{center}
\caption{Typical behavior of internal energy energy in terms of $T$ for $N=50$ and we set unit values for all parameters. Blue dotted line represents the case of $\epsilon=k_l=0$, green dashed line represents the case of $\epsilon=1$, $k_l=0$, orange dash dotted line represents the case of $\epsilon=1$, $k_l=0.6$, red solid line represents the case of $\epsilon=1$, $k_l=1$.}
 \label{fig2}
\end{figure}

In this section,  we review the clustering of galaxy and exact equation
of states in brane world corrected  Newtonian potential \cite{mirh}. It is possible to consider super-light modes in a
brane world models, and they modify the large distance behavior of the Newton's law as
\begin{equation}
\Phi_{i,j}=-\frac{Gm^2}{(r_{ij}^2+\epsilon^2)^{1/2}}\biggl(1+\frac{k_l}{(r_{ij}^2+\epsilon^2)}\biggr),\label{pot}
\end{equation}
where relative position vector (between $i$ and $j$ particles) is
$r_{ij}>>\Lambda={|k_l|}^{1/2}$, where $\Lambda$ is considered as a typical length scale at which correction due to these super-light modes becomes dominant. The parameter $\epsilon$ is a regularization parameter, which occurs due to the extended structure of galaxies. The reason 
for considering extended structure is as following.
It is clear from expression (\ref{ori}) that  the  potential energy  diverges
for the point-mass    (i.e., $r=0$) nature of galaxies. This will lead  to a divergence in the Hamiltonian and, consequently, to the partition function. 
This divergence can be removed by considering  extended nature of galaxies
 (galaxies with halos) with the help of 
the softening parameter $\epsilon$, which assures that the  galaxies are of finite  size \cite{ahm02,hamm}.
The    typical range of the  softening parameter  is $0.01\leq \epsilon\leq 0.05$ in units of the constant cell.   
It may be noted that at small enough distances this modified Newtonian potential reduces to the usual Newtonian potential. This is the limit in which the
contribution from these super-light modes can be neglected. Furthermore, this is required from the physical constraints, as the Newtonian limit of general relativity has been well tested at such scales.
Now it is possible to obtain the two-particle function form this  modified potential as
\begin{equation}
f_{i,j}=\exp{\biggl[\frac{Gm^2}{T(r_{ij}^2+\epsilon^2)^{1/2}}\biggl(1+\frac{k_l}{(r_{ij}^2+\epsilon^2)}\biggr)\biggr]}-1.
\end{equation}
This will further lead to the modification of configurational integrals $Q_N$. For instance, the configurational integral for $N=1$, $Q_{1}(T,V)=V,$
and (for large $r$ where the higher terms of $\frac{k_l}{(r^2+\epsilon^2)^{1/2}}$ can be neglected) the configurational integral for $N=2$,
\begin{equation}
Q_{2}(T,V)=4\pi V \int_{0}^{R_{1}}\left[r^2+\left(\frac{Gm^{2}}{T}\right) \frac{r^{2}}{(r^{2}+\epsilon^{2})^{1/2}}\left(1+ \frac{k_l}{(r^2+\epsilon^2)}\right)\right]dr.
\end{equation}
Evaluating the integrals, we  obtain
\begin{eqnarray}
Q_{2}(T,V)=V^2\left(1+  \alpha_1 x +\alpha_2 x\right),
\end{eqnarray}
where $x=\frac{3}{2} G^3m^6 \bar\rho T^{-3}$ and
\begin{eqnarray}
\alpha_1 &=& \sqrt{1+\frac{\epsilon^2}{R_1^2}}+\frac{\epsilon^2}{R_1^2}\log \frac{\epsilon}{R_1+\sqrt{R_1^2+\epsilon^2}},\\
\alpha_2  &=& -2\frac{k_l}{R_1\sqrt{R_1^2+\epsilon^2}}+2\frac{k_l}{R_1^2}\log \frac{R_1+\sqrt{R_1^2+\epsilon^2}}{\epsilon}.
\end{eqnarray}
Now, we write the most general form  for the  configurational integrals after including modified potential energy  as
\begin{eqnarray}
Q_{N}(T,V)&=& V^N\big(1+ \alpha_1 x +\alpha_2 x\big)^{N-1},\nonumber \\
&=& V^N\big(1+A x\big)^{N-1},\,
\end{eqnarray}
where we have defined $A  = \alpha_1  + \alpha_2$, and these are in turn obtained by solving the configuration integral.
Using this expression, the partition function for the system of galaxies can be written as
\begin{equation}
Z_N(T,V)=\frac{1}{N!}\left(\frac{2\pi mT}{\lambda^2}\right)^{3N/2}V^{N}\big(1+A x\big)^{N-1}.
\end{equation}
Here $\lambda$ refers the normalization factor resulting from
integration over momentum space.
It may be noted that this partition function is expressed as a sum over different order terms, and for a given system, we can restrict this to a certain order of accuracy.\\
We find that the partition function in increasing function of the super-light parameter, however at low temperature, variation with $k_{l}$ is infinitesimal. Furthermore, as this is the partition function  of a system for galaxies interacting through a modified Newtonian potential, we can use it to analyze the effect of super-light modes on the thermodynamics of this system.\\
Thus, the Helmholtz free energy of a system of galaxies corrected by super-light modes in a brane world model can be written as
\begin{equation}
F=-T \ln\biggl(\frac{1}{N!}\big(\frac{2\pi mT}{\lambda^2}\big)^{3N/2}V^N\big(1+A x\big)^{N-1}\biggr).
\end{equation}
In the Fig. \ref{fig0}  we can see typical behavior of Helmholtz free energy in terms of $N$ by variation of $k_{l}$. We find that it is decreasing function of $k_{l}$ in negative region, which means increasing net value of Helmholtz free energy under brane world parameter. There is special case where cases of $\epsilon=k_l=0$ and $\epsilon=1$, $k_l=1.4$ yields to the same Helmholtz free energy.
Now, let us consider a large number of galaxies, i.e., we take a  large value of $N$, and use  $N-1\approx N$. As the inter-galactic distance are very large,  an  collision of galaxies will not usually occur, and can be neglected. Thus, we can write the  entropy $S$ of a system of galaxies corrected by super-light modes as

\begin{equation}
S=N\ln(\bar\rho^{-1}T^{3/2})+N\ln\big(1+A x\big)-3NB+S_{0},
\end{equation}
where $S_{0}=\frac{5}{2}+\frac{3}{2} \ln\left(\frac{2\pi m}{\lambda^2}\right)$ and clustering parameter  is
\begin{equation}
B =\frac{A x}{1+A x}.
\end{equation}
Thus, the clustering parameter depends on $A$, which is obtained by solving the configurational integral for this system
corrected by super-light brane modes.
The internal energy $U$, which is   defined as $U =  F+TS$, can be written as
\begin{equation}
U=\frac{3}{2}NT\big(1-2B\big).
\end{equation}
It may be noted that this internal energy depends on the clustering parameter, which in turn depends on $A$, and that is obtained as  a solution to the configurational integral for a system corrected by these super-light modes.
Thus, the internal energy of this system will also depend on these super-light modes. In the Fig. \ref{fig2} we can see that internal energy is decreasing function of $k_{l}$. We can also see that there is a minimum with negative value of $U$. However, at high temperature and low temperature limit there is no effect with $k_{l}$.

\begin{figure} 
\begin{center}$
\begin{array}{cccc}
\includegraphics[width=70 mm]{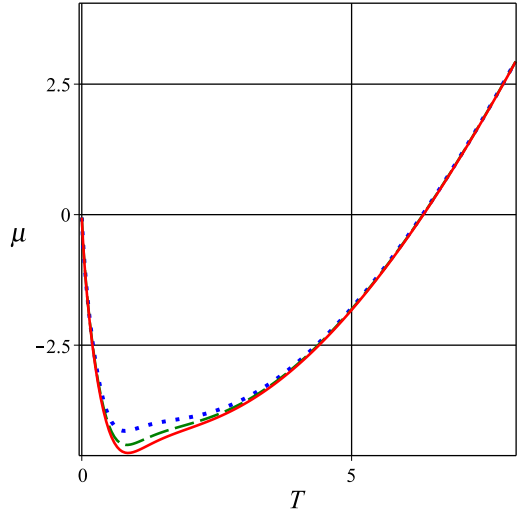}
\end{array}$
\end{center}
\caption{Typical behavior of chemical potential in terms of $T$. We set unit values for all parameters. Blue dotted line represents the case of $k_l=0$, green dashed line represents the case of $k_l=0.6$, red solid line represents the case of $k_l=1$.}
\label{fig3}
\end{figure}

We can also write an expression for the pressure of this system, using the following thermodynamics relation,

\begin{equation}
P= -\biggl(\frac{\partial F}{\partial V}\biggr)_{N,T},
\end{equation}
which yields to the following expression \cite{mirh}
\begin{equation}
P=\frac{NT}{V}\big(1-B\big).
\end{equation}
The chemical potential of this system of galaxies can be expressed as \cite{mirh},
\begin{equation}
\frac{\mu}{T}=\ln(\bar\rho T^{-3/2})-\ln\big(1+A x\big)-\frac{3}{2}\ln\left(\frac{2\pi m}{\lambda^2}\right)-B.
\end{equation}
In the Fig. \ref{fig3} we can see that chemical potential is decreasing function of $k_{l}$. We can also see that there is a minimum with negative value of $\mu$. However, at high temperature and low temperature limit there is no effect with $k_{l}$.

The inclusion of the   super-light modes  modifies the behavior of  $B$, and this in turn modifies the thermodynamics of the system.

\section{Virial Theorem}
In this section, we will analyze the effect of super-light brane modes, on the virial theorem for galaxies. This has been studied for the usual Newtonian potential by analyzing the  adiabatic growth of gravitational clustering \cite{ahm10}. So, for such a system,  the first law of thermodynamics can be written as
\begin{equation}
\frac{d(uR^3)}{dt}+P\frac{dR^3}{dt}=0,
\end{equation}
where $u$ is the energy density of considered matter, $P$ is the pressure and $R(t)$ is the scale factor. Now it is possible to consider a  spherical system of volume $V$ which contains $N$ galaxies. It is possible to associate an thermodynamic temperature $T$,   energy $U$, and pressure $P$  with this system of galaxies. This has been done for a system of galaxies interacting through the usual Newtonian potential \cite{sas00}, and here we obtain this for a system of galaxies interacting thought a potential modifies by super-light brane world modes, and we get \cite{modes}
\begin{eqnarray}
U=\frac{3}{2}NT+\frac{N\bar\rho}{2}\int_{V}\Phi(r)\xi(r)4\pi r^2dr,\\
 P=\frac{NT}{V}-\frac{\bar\rho^2}{6}\int_{V}r\frac{d\Phi(r)}{dr}\xi(r)4\pi r^2dr.
\end{eqnarray}
Now using the explicit expression for the interaction  large distance modified Newtonian potential   between two galaxies (\ref{pot}), we can write  expression for energy and pressure for this system as
\begin{eqnarray}
U&=&\frac{3}{2} NT+W_{N}+W_{M},\label{10}\\
P&=&\frac{3NT+W_N+\epsilon^{2}W_N^{\prime}+3W_M+3\epsilon^{2}W_M^\prime}{3V},
\label{p}
\end{eqnarray}
where the large distance corrected   gravitational
correlation energies can be written as
\begin{eqnarray}
W_N&=&\frac{GN\bar\rho m^2}{2}\int_V\frac{\xi(r)}{(r^2+\epsilon^2)^{\frac{1}{2}}}4\pi r^2dr,
 \nonumber\\
W_N^{\prime}&=&-\frac{GN\bar\rho m^2}{2}\int_V\frac{\xi(r)}{(r^2+\epsilon^2)^{\frac{3}{2}}}4\pi r^2dr,
  \nonumber\\
W_M&=&k_{l}\frac{GN\bar\rho m^2}{2}\int_V\frac{\xi(r)}{(r^2+\epsilon^2)^{\frac{3}{2}}}4\pi r^2dr,
  \nonumber\\
W_M^{\prime}&=&-k_{l}\frac{GN\bar\rho m^2}{2}\int_V\frac{\xi(r)}{(r^2+\epsilon^2)^{\frac{5}{2}}}4\pi r^2dr.
\end{eqnarray}
As the total energy for a spherical volume of radius $R$ can be written as $U=(4/3)\pi uR^3$, and so  from (\ref{10}), we get
\begin{equation}
 \frac{4}{3}\pi uR^3= \frac{3}{2} NT+W_{N}+W_{M} =K+W,
\end{equation}
where $W=W_N+W_M$.
Exploiting  (\ref{p}),  the relation $\frac{d}{dt}R^3=\frac{\dot R}{R}(3R^3)$
reduces to
\begin{equation}
\frac{d}{dt}(R^3)=\frac{\dot R}{R}\bigg[\frac{3}{4\pi}\frac{2K+W_N+\epsilon^{2}W_{N}^{\prime}+3W_M+(l+1)\epsilon^{2}W_M^{\prime}}{P}\bigg],
\end{equation}
where we have used $V=\frac{4}{3}\pi R^3$.

The law of conservation of energy for this system of galaxies can be written as
\begin{equation}\label{15}
\frac{d}{dt}(K+W)+\frac{\dot R}{R}(2K+W_N+\epsilon^{2}W_{N}^{\prime}+3W_M+3\epsilon^{2}W_M^{\prime})=0.
\end{equation}
In the limit  case, when we neglect the  expansion, i.e., $\dot R=0$, we obtain
\begin{equation}
2K+W_N+\epsilon^{2}W_{N}^{\prime}+3W_M+3\epsilon^{2}W_M^{\prime}=0.
\end{equation}
This is the virial theorem for the system of galaxies interacting by  the large distance modified  Newtonian
potential. It can be used to understand the effects of super-light brane modes on the virial theorem. The  general equation for  the law of conservation of energy is given by Eq.  (\ref{15}), and it can lead interesting  limiting cases.
This equation for an extended structures with Newtonian potential can be written as
\begin{equation}
\frac{d}{dt}(K+W)+\frac{\dot R}{R}(2K+W_N+\epsilon^{2}W_{N}^{\prime})=0.
\end{equation}
Now, if we neglect the effects of an extended structure of galaxies, i.e $\epsilon=0$, and so in this limit, a system interacting by the  Newtonian potential, can be described by \cite{ahm10}
\begin{equation}
\frac{d}{dt}(K+W)+\frac{\dot R}{R}(2K+W_N)=0.
\end{equation}
By neglecting the extended structure of galaxies, and only considering the modification by super-light brane modes, we obtain,
 \begin{equation}
\frac{d}{dt}(K+W)+\frac{\dot R}{R}(2K+W_N+3W_M)=0.
\end{equation}
Thus,  we have analyzed the virial theorem for a system of galaxies using the large distance modification to the   Newtonian potential due to super-light modes in the brane world models.

\section{Validity of Extensivity}
In this paper we have used  extensivity, and this approximation is valid for   infinite systems whose thermodynamic functions are  extensive. An  requirement for such an approximations is that the correlation energy between the cells should  be less than the correlation energy within an average cell \cite{sas96}. We assume the size of the cells is much larger than the correlation length, and a power law behavior for the two-point correlation function, which is given by \cite{sas80}
\begin{equation}
\xi(r)=\xi_0r^{-\gamma},
\end{equation}
where $\gamma \sim 1.8$  is a constant parameter.

The correlation energy for an extended mass in a spherical volume $V$ is given by
\begin{equation}
W_M(V)=-\frac{Gm^2\bar\rho^2V}{2}I_{1},
\end{equation}
where
\begin{equation}
I_{1}=\int_{0}^{R_1}\biggl[\frac{1}{(r^2+\epsilon^2)^{1/2}}\biggl(1+\frac{k_l}{(r^2+\epsilon^2)}\biggr)\biggr]\frac{\xi_0}{r^\gamma}4\pi r^2dr.
\end{equation}
Thus, for a system of galaxies interacting by a Newtonian potential modified by super-light brane modes, we obtain
\begin{equation}
W_M(V)=W_N(V)\biggl[1+\biggl(\frac{2-\gamma}{2\gamma}+\frac{2\gamma}{\gamma+2}\frac{k_l}{R_{1}^2}(3)\biggl)\frac{\epsilon^2}{R_{1}^2}+\biggl(\frac{2-\gamma}{-\gamma}\frac{k_l}{R_{1}^2}\biggl)\biggr], \label{w}
\end{equation}
where
\begin{equation}
W_N(V)=-2\pi Gm^2\bar\rho^2V\xi_0\frac{R_1^{2-\gamma}}{2-\gamma}.
\end{equation}
Now using  $2V$, this expression for $W_N$ can  be written as
\begin{equation}
W_N(2V)=-2\pi Gm^2\bar\rho^2(2V)\xi_0\frac{(2^{1/3}R_1)^{2-\gamma}}{2-\gamma}.
\end{equation}
Thus, we can write $W_M$ as
\begin{equation}\label{2v}
W_M(2V)=2(2^{1/3})^{2-\gamma}W_N(V)I_{2},
\end{equation}
where
\begin{eqnarray}\label{2v-2v}
I_{2}&=&\biggl[1+\biggl(\frac{2-\gamma}{2\gamma}+\frac{2\gamma}{\gamma+2}\frac{k_l}{(2^{1/3}R_{1})^2}(3)\biggl)\frac{\epsilon^2}{(2^{1/3}R_{1})^2}\nonumber\\
&+&\biggl(\frac{2-\gamma}{-\gamma}\frac{k_l}{(2^{1/3}R_{1})^2}\biggl)\biggr].
\end{eqnarray}
So, the correlation energy for an extended mass in a spherical volume   (\ref{w}),
can be expressed as
\begin{equation}
W_M(V)=W_N(V)\biggl[1+f(\gamma,2,\frac{\epsilon}{R_{1}})\biggr],
\end{equation}
where
\begin{equation}
f(\gamma,2,\frac{\epsilon}{R_{1}})=\biggl(\frac{2-\gamma}{2\gamma}+\frac{2\gamma}{\gamma+2}\frac{k_l}{R_{1}^2}(3)\biggl)\frac{\epsilon^2}{R_{1}^2}+\biggl(\frac{2-\gamma}{-\gamma}\frac{k_l}{R_{1}^2}\biggl).
\end{equation}
However,   the correlation energy for an extended mass in $2V$ can   be written      as
\begin{equation}
W_M(2V)=2(2^{1/2})^{2-\gamma}W_N(V)\biggl[1+h(\gamma,2,\frac{\epsilon}{R_{1}})\biggr],
\end{equation}
with
\begin{eqnarray}
h(\gamma,2,\frac{\epsilon}{R_{1}})&=&\biggl(\frac{2-\gamma}{2\gamma}+\frac{\gamma}{\gamma+2}\frac{6k_l}{(2^{1/3}R_{1})^2}\biggl)\frac{\epsilon^2}{(2^{1/3}R_{1})^2}\nonumber\\
&-&\biggl(\frac{2-\gamma}{\gamma}\frac{k_l}{(2^{1/3}R_{1})^2}\biggl).
\end{eqnarray}
For extensivity to be a valid  approximation, we should have  $|W_{2V}/W_V|\sim 1$.
Thus, we obtain
\begin{equation}
\biggl|{\frac{W_{M2V}}{2W_{MV}}}\biggr|=(2^{1/3})^{2-\gamma}g(\gamma,2,\epsilon/R_{1}),
\end{equation}
where
\begin{equation}
g(\gamma,l,\epsilon/R_{1})=\frac{1+f(\gamma,2,\epsilon/R_{1})}{1+h(\gamma,2,\epsilon/R_{1})}.
\end{equation}
Here, we observe that the validity of the extensivity approximation   depends upon the values of $\gamma$.

\section{Cosmic energy equation}
The cosmic energy equation has been studied for a system of galaxies interacting through a Newtonian potential \cite{sas86,sut90}.
 In this section, we will analyze the effect of super-light modes on the cosmic energy equation. Thus, we can start from the expression
\begin{equation}
\frac{d}{dt}(K+W)+\frac{\dot R}{R}(2K+W_N+\epsilon^{2}W_{N}^{\prime}+3W_M+3\epsilon^{2}W_M^{\prime})=0.
\end{equation}
This can be written as
\begin{equation}
\frac{d}{dt}(K+W)+\frac{\dot{R}}{R}(2K+W_{N}+\eta W_{N})=0,\label{df}
\end{equation}
where $\eta$ is a constant defined by
\begin{equation}
\eta=\epsilon^{2}\frac{W_{N}^{\prime}}{W_{N}}+3\frac{W_{M}}{W_{N}}+3\epsilon^2\frac{W_{M}^{\prime}}{W_{N}}.
\end{equation}
The ratio of gravitational
correlation energy  to the kinetic energy, for this system of interacting extended mass galaxies can be written as
\begin{equation}
B=-\frac{W}{2K}.
\end{equation}
This has been analyzed for a system of galaxies interacting thought a usual Newtonian potential \cite{sas86,sut90},  and we will analyze the correction to that from super-light brane modes. Thus, for a system of galaxies interacting thought a large distance modified brane world Newtonian potential, we obtain
\begin{equation}
\frac{dy_{1}}{dt}-\frac{(2-y_{1})}{W }\frac{dW }{dt}-2\frac{\dot{R}}{R}(1-y_{1}+\eta)=0,
\label{55}
\end{equation}
where $y_{1}= 1/B$.
Now, using a relation
\begin{equation}
y_{1}=\frac{y+A-1}{A},
\end{equation}
where $A  = \alpha_1  + \alpha_2$ and $y=1/B_0$, where $B_0$ is the ratio of
correlation energy  to the kinetic energy for  point mass galaxies.
In fact, these values are obtained by solving the configurational integrals.
The evolution, of this system of galaxies  interacting through large distance   modified Newtonian potential is given by
\begin{equation}
\frac{dy}{dt}-\frac{(1+A-y)}{W}\frac{dW}{dt}-2\frac{\dot{R}}{R}(1-y+A\eta)=0.
\label{ki}
\end{equation}
The expansion factor $R(t)$ depends on time $t$ as
\begin{equation}
R(t)\propto t^{s}. \label{r}
\end{equation}
This  is a general expression valid for any cosmological model  for a  constant  $s$.
 Now, we can write $W_{N}+ W_M$  as
 \begin{equation}
 W(t)\propto t^{\beta}.\label{rt}
 \end{equation}
 This equation can be obtained by using   $\frac{d}{dt}=\dot R\frac{d}{dR}$, and the  BBGKY hierarchy  method  \cite{ina76}. Here we have considered   $(
 \bar\rho\propto R^{-3})$.
  Thus, we can use such a power law dependence for $R(t)$ and $W(t)$
to solve (\ref{ki}).
Using  (\ref{w}) and (\ref{rt}), Eq. (\ref{ki}) reduces to
\begin{equation}
\frac{dy(t)}{dt}=\frac{(1+A-y)\beta}{t}+\frac{2s(1-y+A\eta)}{t},
\end{equation}
where
 $s=1/2$, $2/3$ and $1$ corresponds to Dirac, Einstein-de Sitter and Milne Universes, respectively.
 The above equation can be solved for $y(t)$, and so we obtain
\begin{eqnarray}
 y(t)= y_{c}+ (y_{0}-y_{c}) t^{-\beta -2 s},
\end{eqnarray}
 where $y_{0}$ corresponds to $y$ at $t=0$.
The asymptotic value (or critical value) of $y(t)$ is given by:
\begin{equation}
 {y_{c}}=\frac{1}{
B_{c}}=\frac{A(\beta +2 \eta  s)+\beta +2 s}{\beta +2 s}.\label{bc}
\end{equation}
This has been done for a system of galaxies interacting through a usual potential \cite{sut90} , and here we will analyze the effects of large distance corrected brane world Newtonian potential on this system.  Thus, again  $B=B_{c}$ will  not depend on the present value $B$, and can be expressed in terms of  $\beta$. The value of $\beta$ is sensitive to the primordial power spectrum law, i.e.,  $n$ , and is roughly fitted, by simulations, to
\begin{equation}
\beta\sim\frac{1-n}{3}.
\end{equation}
For this value of $\beta$, $B_{c}$ in (\ref{bc}) becomes,
\begin{equation}
B_{c}\sim \frac{(1-n)+6s}{A(1-n+6\eta s) +1-n +6s}.
\end{equation}
So,    the value of $B_{c}$  depends on $A$, but the value of $A$ depends on the super-light brane modes. Thus, the super-light brane modes would modify the clustering properties of a system of galaxies interacting.

\section{Conclusion}
In this paper, we have analyzed the thermodynamics of gravitational clustering of galaxies in brane world models.
We have analyzed the modification to the Newtonian potential produced by super-light brane modes. We have used   an adiabatic approximation for  performing this  analysis. We calculated partition function and found that is increasing function of super-light parameter, analyzing of Helmholtz free energy shows that is decreasing function of super-light parameter and yields to negative infinity for the large super-light parameter. We also found that the internal energy as well as chemical potential is decreasing function of super-light parameter. We observed that the clustering parameter  gets modified in  the  brane world models.  We calculate the corrections to the
virial theorem from brane world models.
Further, we have discussed the correction to the validity of extensivity,
which is influenced by $\gamma$. The corrections to the clustering parameter modify the cosmic energy equation. We also solved   the corrected cosmic energy equation and analyzed the asymptotic behavior.  It was observed that the large scale modification of Newtonian potential directly affects the solutions of the cosmic energy equation.\\

The modification to the Newtonian potential considered in this paper,   occurred due
to the super-light  modes in the  brane world models. It may be noted that the Newtonian potential also gets modified from various different approaches. These include non-commutative geometry \cite{nic10,gre13},   minimal length in quantum gravity \cite{ali13},    f(R)  gravity  \cite{noj08,sud11}, dark energy \cite{sud} and  the  entropic force  \cite{maj13}. It would be possible to analyze super-light modes in such deformed theories, and this will have an effect on the large scale behavior of Newtonian potential. Thus, we can use such a large scale corrected Newtonian potential, and analyze its effects on clustering of galaxies. So, it would be interesting to  analyze the clustering using these modifications to the Newtonian potential. It would
also be interesting to analyze the effect of these  modifications to the Newtonian potential on the cosmic energy equation. It is expected that the
virial theorem  will get corrected due to the deformation of the Newtonian potential. The correction to the virial theorem will effect the  cosmic energy equation.

\end{document}